\documentstyle[aps,epsf,multicol,amsmath,amsfonts,graphicx,bbm]{revtex}

\begin{document}

\title{Lossless quantum data compression and variable-length coding}

\author{Kim Bostr\"om and Timo Felbinger}
\address{
}
\date{\today}
\maketitle

\begin{abstract}

In order to compress quantum messages without loss of information it is necessary to allow the length of the encoded messages to vary. We develop a general framework for variable-length quantum messages in close analogy to the classical case and show that lossless compression is only possible if the message to be compressed is known to the sender. The lossless compression of an ensemble of messages is bounded from below by its von-Neumann entropy. We show that it is possible to reduce the number of qbits passing through a quantum channel even below the von-Neumann entropy by adding a classical side-channel. We give an explicit communication protocol that realizes lossless and instantaneous quantum data compression and apply it to a simple example. This protocol can be used for both online quantum communication and storage of quantum data.

\end{abstract}

\begin{multicols}{2}
\narrowtext

\section{Introduction}

Any physical system can be considered as a carrier of information because the state of that system could in principle have been intentionally manipulated to represent a \emph{message}. The state of a system composed from distinguishable subsystems forms a message of a certain \emph{length}, where each subsystem represents one \emph{letter}. In quantum information theory, the systems are quantum and the system states represent quantum messages. A message is compressed if it is mapped to a shorter message and if this map is reversible, then no information has been lost. 
\emph{Schumacher} was the first to present a method for quantum data compression \cite{Schumacher95}. It is
based on the concept of encoding only a \emph{typical subspace} spanned by the typical sequences emitted by a memoryless source. Since then there have been further investigations \cite{Schumacher94,Schumacher96,Barnum96,Jozsa98,Braunstein98,Chuang00,Schumacher00}, but all considered compression methods are only faithful in the limit of large block lengths. 
Now we ask: Is it possible to compress quantum messages without \emph{any} loss of information? To answer this question some basic concepts of quantum information theory have to be revisited. In particular, the requirement of a fixed block length for quantum messages has to be abandoned and must be replaced by a more general theory of quantum messages which enables a flexible and easy treatment of quantum codes involving codewords of variable-length. At first, we develop a general framework in close analogy to the classical case, based on previous work by one of us \cite{Bostroem00mlconcepts,Bostroem00mlcoding}. A different approach to variable-length quantum messages (appearing as a special case in our formalism) has been worked out by \emph{Braunstein et al.} \cite{Braunstein98} and \emph{Schumacher and Westmoreland} \cite{Schumacher00}. 
We define a measure of information quantifying the effort of communication. Compression then means reducing this effort. We argue that prefix codes are practically not very useful for quantum coding and suggest a different method involving an additional classical side-channel. With the help of this channel, certain problems of instantaneous quantum communication can be avoided and, moreover, the quantum channel can be used with higher efficiency. At last, we present a communication protocol that enables lossless and instantaneous quantum data compression and we demonstrate its efficiency by an explicit example.
Let us start with reviewing the fundamental notion of a \emph{code}\footnote{Some notions and definitions are already existing, some are based on our own reasoning. When we find an already existing definition equal or similiar to the desired one, we use it and in case it is not a standard definition, we give an explicit reference. For a profound review on classical information theory, see \cite{CoverThomas91,MacKay}, for a profound review on quantum information theory, see \cite{NielsenChuang01,Preskill}.}.
\begin{figure}\vspace{-2em}
	\[{\includegraphics[width=0.5\textwidth]{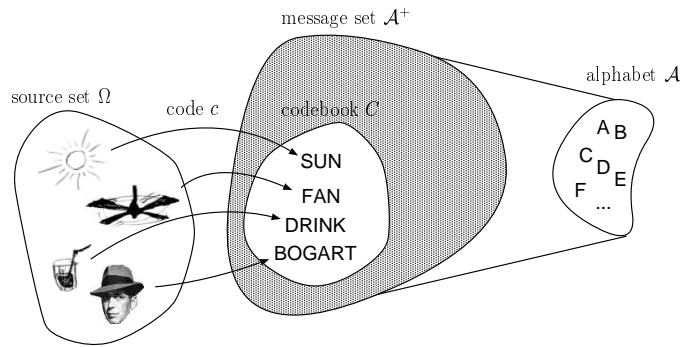}}\]
	\vspace*{-0.5cm}\caption{\small A classical code is a map from a set of source objects into a set of codewords composed from an alphabet. An ensemble of source objects is mapped to an ensemble of codewords. For variable-length codes, the length of the codewords is allowed to vary.}\label{ccode}
\end{figure}

\section{Codes}

Basically, when you have a set of things and you want to give them a name, then this is a \emph{coding} task. There is a code for bank accounts, telephone devices and inhabitants of a country, there even is a code for living beings: the genetic code. Language is a code for thoughts, which are in turn codes for abstract ideas or concrete objects of human experience. 
A code gives \emph{meaning} to a message, it relates objects to their description.
Objects are encoded into messages composed from a basic \emph{alphabet}. The number of letters that is needed to describe a particular object is a good measure of the \emph{information content} given to the object by the code. This is the key to \emph{data compression} which we will study in the following with a focus on quantum codes.

Classically, a \emph{code} is a map $c:\Omega\rightarrow M$ from a set of objects, $\Omega$, to a set of messages, $M$ (see Fig.~\ref{ccode}). It is the messages that can be communicated and not the objects themselves, so communication is always based on a code. \emph{Messages} (or \emph{strings}) are sequences of letters taken from an \emph{alphabet} ${\cal A}$ and are denoted by $x^n:=x_1\cdots x_n,\quad x_i\in{\cal A}$. The \emph{empty message} is denoted by $x^0:=\o$. 
All messages of length $n$ form the set 
\begin{equation}\label{cn}
	{\cal A}^n :=\{x^n\mid x_i\in{\cal A}\},
\end{equation}
and the empty message forms the set ${\cal A}^0:=\{\o\}$. 
All strings of finite length form the set of \emph{general messages} over the alphabet ${\cal A}$,
\begin{equation}\label{cplus}
	{\cal A}^+:=\bigcup_{n=0}^\infty {\cal A}^n.
\end{equation}
Every subset $M\subset{\cal A}^+$ is a \emph{message set}.
Now we can precisely define a classical \emph{$k$-ary code} as a map $c:\Omega\rightarrow{\cal A}^+$ with $k:=|{\cal A}|$. The set $ C=c(\Omega)$ is the \emph{codebook} and each member of $ C$ is a \emph{codeword}. 
Being a subset of ${\cal A}^+$, a codebook is also a message set (just like a nightingale is also a bird).
If $ C\subset{\cal A}^n$ for some $n\in{\mathbbm N}$, then $c$ is called a \emph{block code}, otherwise a \emph{variable-length code}. 
There is another important classification: \emph{lossless} and \emph{lossy} codes. A code is \emph{lossless} (or \emph{uniquely decodable} or \emph{non-singular}), if there are distinct codewords for distinct objects, i.e. \mbox{$\forall x,y\in\Omega:\ x\neq y\Rightarrow c(x)\neq c(y)$}. In case of a \emph{lossy} code, some objects are mapped to the same encoding.
Lossy codes are used when it is more important to reduce the size of the message than to ensure the correct decoding (a fine example is the MP3 code for sound data). For a given probability distribution on $\Omega$, lossy codes can also be useful if the \emph{fidelity} $F$, i.e. the probability of correct decoding, is close to 1. For lossless codes the fidelity is exactly 1. In this paper, we only consider lossless codes.

\subsection{The general message space}

The transition from classical to quantum information is simple. We just allow the elements of a source set $\Omega$ to be in \emph{superposition}. Precisely, we interpret $\Omega$ as an orthonormal basis for a Hilbert space ${\cal V}$ and consider every normalized vector of ${\cal V}$ as
a valid object. Then ${\cal V}$ is the \emph{linear span} of $\Omega$ and we write ${\cal V}={\rm Span}(\Omega)$ with $\dim{\cal V}=|\Omega|$.
The same goes for the messages. We interpret a message set $M$ as an orthonormal basis for a message space ${\cal M}={\rm Span}(M)$ with $\dim{\cal M}=|M|$ and consider each element of ${\cal M}$ as a valid message.
The map $c:{\cal V}\rightarrow{\cal M}$ then represents a \emph{quantum code} with the space ${\cal C}=c({\cal V})$ being the \emph{code space} and the elements of ${\cal C}$ being the \emph{codewords}. In order to preserve linearity, the code must be a linear map and in order to preserve norm, the code must be an isometric map.
In the literature, often the code space ${\cal C}$ rather than the map $c$ is called a code (this is a bit like calling $f(x)$ a function). However, by saying ``code'' we will refer to the map $c$ here, in full analogy to the classical case. 
Now let us find the general message space corresponding to the classical general message set ${\cal A}^+$. Interpret the letters of a \emph{quantum alphabet} ${\cal Q}$ as an orthonormal basis for a \emph{letter space} ${\cal H}:={\rm Span}({\cal Q})$.
A letter space ${\cal H}$ with $k=\dim{\cal H}=|{\cal Q}|$ is called a \emph{$k$-ary space}.
Quantum letters are composed into messages by tensor multiplication, giving \emph{product messages} $|x^n\rangle:=|x_1\rangle\otimes\cdots\otimes|x_n\rangle$ that form the set ${\cal Q}^n:=\{|x^n\rangle\mid |x_i\rangle\in{\cal Q}\}$ and span the \emph{block space} ${\cal H}^{\otimes n}:={\rm Span}({\cal Q}^n)$, giving
\begin{equation}
	{\cal H}^{\otimes n}=\bigotimes_{i=1}^n {\cal H}={\cal H}\otimes\cdots\otimes{\cal H}.
\end{equation}
The space ${\cal H}^{\otimes n}$ is the quantum analogue to the set ${\cal A}^n$ of classical block messages given by~(\ref{cn}), and contains arbitrary superpositions of product messages, which are called \emph{entangled messages}. Because superposition and entanglement have no classical interpretation, quantum information is truly different from classical information.
The empty message, denoted by $|x^0\rangle\equiv|\o\rangle$, forms the set ${\cal Q}^0=\{|\o\rangle\}$ and spans the one-dimensional space ${\cal H}^{\otimes 0}:={\rm Span}({\cal Q}^0)$. Elements of ${\cal H}^{\otimes n}$ for some $n\in{\mathbbm N}$ are called \emph{block messages}.
The set of all product messages composed from ${\cal Q}$ is denoted by ${\cal Q}^+:=\bigcup_{n=0}^\infty{\cal Q}^n$. Now the \emph{general message space} ${\cal H}^{\oplus}$ induced by ${\cal H}$ can be defined by ${\cal H}^{\oplus}:={\rm Span}({\cal Q}^+)$, giving
\begin{equation}
	{\cal H}^{\oplus}=\bigoplus_{n=0}^\infty{\cal H}^{\otimes n}
	={\cal H}^{\otimes 0}\oplus{\cal H}\oplus{\cal H}^{\otimes 2}\oplus\cdots.
\end{equation}
The space ${\cal H}^{\oplus}$ is the quantum analogue to the set ${\cal A}^+$ of general classical messages given by~(\ref{cplus}).
${\cal H}^{\oplus}$ is a separable Hilbert space with the countable basis ${\cal Q}^+$.
The space ${\cal H}^{\oplus}$ is similiar to the Fock space in many-particle theory, except that the particles are letters here, which must be distinguishable, so there is no symmetrization or antisymmetrization.
The general message space contains also superpositions of messages of distinct length, for example
\begin{equation}
	\frac1{\sqrt2}(|101\rangle+|11100\rangle)\in{\cal H}^{\oplus},
\end{equation}
if $|0\rangle,|1\rangle\in{\cal H}$. Any block space ${\cal H}^{\otimes n}$ is a subspace of ${\cal H}^{\oplus}$ and is orthogonal to any other block space ${\cal H}^{\otimes m}$ with $n\neq m$. Elements with components of distinct length are called \emph{variable-length messages} (or \emph{indeterminate-length messages}) to distinguish them from block messages. Any subspace ${\cal M}\subset{\cal H}^{\oplus}$ is called a \emph{message space} and its elements are \emph{quantum messages}.

\subsection{Length operator}
\label{sec:lengthop}

Define the \emph{length operator} in ${\cal H}^{\oplus}$ measuring the length of a message as
\begin{equation}\label{lengthop}
	\hat L:=\sum_{n=0}^\infty  n\,\Pi_ n,
\end{equation}
where $\Pi_ n$ is the projector on the block space ${\cal H}^{\otimes  n}\subset{\cal H}^{\oplus}$, given by
\begin{equation}
	\Pi_ n=\sum_{ x^ n\in{\cal Q}^n}| x^ n\rangle\langle x^ n|.
\end{equation}
As $\hat L$ is a quantum observable, the length of a message $| x\rangle\in{\cal H}^{\oplus}$ is generally not sharply defined. Rather, the measurement of $\hat L$ generally disturbs the message by projecting it on a block space of the corresponding length. The \emph{expected length} of a message $| x\rangle\in{\cal H}^{\oplus}$ is given by 
\begin{equation}\label{exp_length}
	 L( x):=\langle x|\hat L| x\rangle. 
\end{equation}
However, in ${\cal H}^{\oplus}$ there are also messages whose expected length is infinite. Classical analoga are probability distributions with non-existing moments, e.g. the Lorentz distribution.
Block messages are eigenvectors of $\hat L$, that is, $\hat L| x\rangle=n\,| x\rangle$ for all $| x\rangle\in{\cal H}^{\otimes n}$.

The generalization to statistical ensembles is straightforward. Consider an ensemble $\Sigma=\{p,{\cal X}\}$ of variable-length messages $| x\rangle\in {\cal X}\subset{\cal H}^{\oplus}$ occurring with probability $p( x)>0\ \forall| x\rangle\in{\cal X}$ such that $\sum_{ x\in{\cal X}}p( x)=1$. Then there is a density operator
\begin{equation}
	\sigma=\sum_{ x\in{\cal X}}p( x)| x\rangle
	\langle x|,
\end{equation}
called a \emph{statistical quantum message}, representing the ensemble $\Sigma$. The set of all such density operators is denoted by ${\cal S}({\cal H}^{\oplus})$. \emph{Vice versa}, however, for a given density operator $\sigma\in{\cal S}({\cal H}^{\oplus})$ there is in general a non-countable set of corresponding ensembles. In terms of information theory, $\sigma$ cannot be regarded as a lossless code for the ensemble $\Sigma$. There is more information in the ensemble than in the corresponding density operator. 
As we will see, this additional \emph{a priori} knowledge is in fact needed to make lossless compression possible.

The expected length of an ensemble $\Sigma$ or of the corresponding statistical message $\sigma\in{\cal S}({\cal H}^{\oplus})$ is defined as
\begin{equation}
	 L(\Sigma)= L(\sigma):={\rm Tr}\{\sigma\,\hat L\}
	=\sum_{ x\in{\cal X}}p( x)\, L( x).
\end{equation}

\subsection{Base length}

The expected length of a quantum message $| x\rangle$, given by~(\ref{exp_length}), will in general not be the outcome of a length measurement. Every length measurement results in one of the length eigenvalues supported by $| x\rangle$ and generally disturbs the message. If there is a maximum value resulting from a length measurement of a state $| x\rangle$, namely the length of the longest component of $| x\rangle$, then let us call it the \emph{base length} of $| x\rangle$, defined as
\begin{equation}\label{base_length}
	\underline L( x):=\max\{ n\in{\mathbbm N}\mid \langle x|\Pi_ n| x\rangle>0\}.
\end{equation}
For example, the quantum message
\begin{equation}
	| x\rangle=\frac1{\sqrt2}(|abra\rangle+|cadabra\rangle)
\end{equation}
has base length 7. 
Since the base length of a state is the size of its longest component, we have
\begin{equation}
	\underline L( x)\geq  L( x).
\end{equation}
It is important to note that the base length is not an observable.
It is only available if the message $| x\rangle$ is \emph{a priori} known.

\subsection{Quantum code}

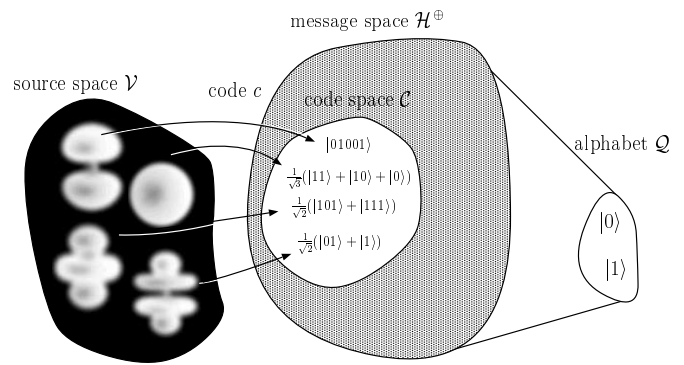
\begin{figure}\vspace{-2em}
	\[{\includegraphics[width=0.5\textwidth]{images/qcode.eps}}\]
	\vspace*{-0.5cm}\caption{\small A quantum code is a linear isometric map from a source space of quantum objects into a code space of codewords composed from a quantum alphabet. Superpositions of source objects are encoded into superpositions of codewords. An ensemble of source objects is mapped to an ensemble of codewords. For a variable-length quantum code, the length of the codewords is allowed to vary. Superpositions of codewords of distinct length lead to codewords of indeterminate length. The \emph{base length} of a codeword is defined as the length of the longest component.}\label{qcode}
\end{figure}

Now we can precisely define a \emph{$k$-ary quantum code} to be a linear isometric map $c:{\cal V}\rightarrow{\cal H}^{\oplus}$, where ${\cal V}$ is a Hilbert space and ${\cal H}^{\oplus}$ is the general message space induced by a letter space ${\cal H}$ of dimension $k$. The image of ${\cal V}$ under $c$ is the \emph{code space} ${\cal C}=c({\cal V})$ (see Fig.~\ref{qcode}). 
Being a quantum analogue to the codebook, ${\cal C}$ is the space of valid codewords.
The code $c$ is uniquely specified by the transformation rule 
\begin{equation}
	|\omega\rangle\stackrel{c}{\longmapsto} |\gamma\rangle,
\end{equation}
where $|\omega\rangle$ are elements of a fixed orthonormal basis ${\cal B}_{\cal V}$ of ${\cal V}$ and $|\gamma\rangle=|c(\omega)\rangle$ are elements of an orthonormal basis ${\cal B}_{\cal C}$ of ${\cal C}$.
Since $c$ is an \emph{isometric} map, i.e. $\langle\omega|\omega'\rangle=\langle c(\omega)|c(\omega')\rangle$, this implies that $|c(\omega)\rangle\neq|c(\omega')\rangle$ for all $|\omega\rangle\neq|\omega'\rangle$ in ${\cal V}$, so $c$ is a lossless code with an inverse $c^{-1}$. The quantum code $c$ can be represented by the isometric operator
\begin{equation}
	C:=\sum_{\omega\in{\cal B}_{\cal V}}|c(\omega)\rangle\langle\omega|
	=\sum_{\gamma\in{\cal B}_{\cal C}}|\gamma\rangle\langle c^{-1}(\gamma)|,
\end{equation}
called the \emph{encoder} of $c$.
Since $c$ is lossless, there is an inverse operator
\begin{equation}
	D:=C^{-1}=\sum_{\gamma\in{\cal B}_{\cal V}}|\omega\rangle\langle c(\omega)|
	=\sum_{\gamma\in{\cal B}_{\cal C}}|c^{-1}(\gamma)\rangle\langle \gamma|,
\end{equation}
called the \emph{decoder}. 
In practice, the source space ${\cal V}$ and the code space ${\cal C}$ are often subspaces of one and the same physical space ${\cal R}$. Since $C$ is an isometric operator between ${\cal V}$ and ${\cal C}$, there is a (non-unique) \emph{unitary extension} $U_C$ on ${\cal R}$ with
\begin{eqnarray}
	U_C | x\rangle &=& C| x\rangle,
		\quad\forall | x\rangle\in{\cal V}\subset{\cal R},\\
	U_C^\dagger|y\rangle &=& C^{-1}|y\rangle,\quad
		\forall|y\rangle\in{\cal C}\subset{\cal R}.
\end{eqnarray}
However, using $C$ and distinguishing between ${\cal V}$ and ${\cal C}$ is more convenient and more general.
Codes with ${\cal C}\subset{\cal H}^{\otimes n}$ for some $n\in{\mathbbm N}$ are called \emph{block codes}, otherwise \emph{variable-length codes}.

\section{Realizing variable-length messages}

Variable-length messages could in principle directly be realized by a quantum system whose particle number is not conserved, for instance, an electromagnetic field. Each photon may carry letter information by its field mode, while the number of photons may represent the length of the message. The photons can be ordered either using their spacetime position (e.g. single photons running through a wire) or some internal state with many degrees of freedom (e.g. a photon with frequency $\omega_2$ can be defined to ``follow'' a photon with frequency $\omega_1<\omega_2$). The Hilbert space representing such a system of distinguishable particles with non-conserved particle number simply \emph{is} the message space ${\cal H}^{\oplus}$.
In case we have only a system at hand, where the number of particles is conserved, we can also realize variable-length messages by embedding them into block spaces. 

It is a good idea to distinguish between the \emph{message space}, which is a purely abstract space, from its physical realization. Let us call the physical realization of a message space ${\cal M}$ the \emph{operational space} $\tilde{\cal M}$. Between ${\cal M}$ and $\tilde {\cal M}$, there is an isometric map, so $\dim{\cal M}=\dim\tilde{\cal M}$. This is expressed by ${\cal M}\cong\tilde{\cal M}$. The operational space $\tilde{\cal M}$ is the space of physical states of a system representing valid codewords of ${\cal M}$. Often the operational space is a subspace of the total space of all physical states of the system. Denoting the total \emph{physical space} by ${\cal R}$ we have 
\begin{equation}
	{\cal M} \cong \tilde{\cal M}\subset{\cal R}.
\end{equation}

\subsection{Bounded message spaces}\label{realizing}

The general message space ${\cal H}^{\oplus}$ is the ``mother'' of all message spaces induced by the letter space ${\cal H}$. It contains just \emph{every} quantum message that can be composed using letters from ${\cal H}$ and the laws of quantum mechanics. However, it is an \emph{abstract} space, i.e. independent from a particular physical implementation. It would be good to know if such a space can also physically be realized. It is clear that if you have a \emph{finite system} you can only realize a \emph{finite dimensional subspace} of the general message space, whose dimension is infinite. So let us start with the physical realization of the \emph{$r$-bounded message space}
\begin{equation}
	{\cal H}^{\oplus r}:=\bigoplus_{ n=0}^r{\cal H}^{\otimes  n},
\end{equation}
containing all superpositions of messages of maximal length $r$.

Say you have a physical space ${\cal R}={\cal D}^{\otimes s}$ representing a register consisting of $s$ systems with $\dim{\cal D}=k$. Each subspace ${\cal D}$ represents one \emph{quantum digit} in the register. In the case $k=2$ the quantum digits are \emph{quantum bits}, in short ``qbits''.
The physical space ${\cal R}$ represents the space of all \emph{physical states} of the register, while the message space ${\cal H}^{\oplus r}$ represents the space of \emph{valid codewords} that can be held by the register and it is isomorphic to a subspace $\tilde{\cal H}^{\oplus r}$ of the physical space ${\cal R}$. 
Let $\dim{\cal H}=k$, then you must choose $s$ such that
\begin{eqnarray}
	\dim({\cal H}^{\oplus r}) &\leq&\dim({\cal D}^{\otimes s})\\
	\Rightarrow\quad\sum_{n=0}^r k^n=\frac{k^{r+1}-1}{k-1}&\leq& k^s\\
	\Rightarrow\quad s&\geq&r+1.
\end{eqnarray}
Thus you need a register of at least $(r+1)$ digits to realize the message space ${\cal H}^{\oplus r}$. Choose the smallest possible register space ${\cal R}={\cal D}^{\otimes (r+1)}$. Since at most $r$ digits are carrying information, one digit can be used to indicate either the beginning or the end of the message. Now you can conveniently use \emph{$k$-ary representations of natural numbers} as codewords.
Each natural number $i$ has a unique $k$-ary representation $ Z_k(i)$. For instance, $ Z_2(3)=11$ and $ Z_{16}(243)=E3$. All $k$-ary representations have a \emph{neutral prefix} ``0'' that can precede the representation without changing its value, e.g. $000011\cong 11$. 
For a natural number $n>0$, define $ Z_k^n(i)$ as the \emph{$n$-extended $k$-ary representation of $i$} by
\begin{equation}
	 Z_k^n(i) := \underbrace{0\cdots 0 Z_k(i)}_{n},\quad 0\leq i\leq k^r-1.
\end{equation}
For example, $ Z_2^6(3)=000011$ and $ Z_{16}^6(243)=0000E3$.
Let us define that the message starts after the first appearance of ``1'', e.g. $000102540\cong02540$. Now define orthonormal vectors
\begin{equation}\label{nonempty_basis}
	|e_i^n\rangle := |\underbrace{0\cdots 0}_{r-n}1 Z_k^n(i)\rangle
	\in{\cal R}
\end{equation}
where $n>0$ and $0\leq i\leq k^n-1$.
The $n$ digits of $ Z_k^n(i)$ are called \emph{significant digits}.
The empty message corresponds to the unit vector
\begin{equation}
	|\o\rangle := |e_0^0\rangle := |0\cdots 01\rangle.
\end{equation}
Obviously, $|\o\rangle$ has no significant digits.
\begin{figure}[h!]\vspace{-2em}
	\[{\includegraphics[width=0.3\textwidth]{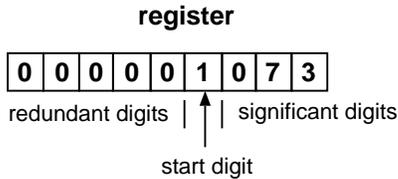}}\]
	\vspace*{-0.5cm}\caption{\small Realizing a general variable-length message.}\label{bild}
\end{figure}
Next, define orthonormal basis sets
\begin{equation}
	\tilde{\cal B}^n := \big\{|e_0^n\rangle,\ldots,|e_{k^n-1}^n\rangle\},\quad
	0\leq n\leq r,
\end{equation}
that span the operational block spaces
\begin{equation}
	\tilde{\cal H}^{\otimes n}={\rm Span}(\tilde{\cal B}^n).
\end{equation}
Note that $\tilde{\cal H}^{\otimes n}$ is truly different from ${\cal H}^{\otimes n}$, because $\tilde{\cal H}^{\otimes n}$ has dimension $k^{r+1}$, while $\tilde{\cal H}^{\otimes n}$ has dimension $k^n$.
Next, define an orthonormal basis
\begin{equation}
	\tilde{\cal B}^{+}:=\bigcup_{n=0}^r\tilde{\cal B}^n,
\end{equation}	
and construct the operational space $\tilde{\cal H}^{\oplus r}\subset{\cal R}$ by
\begin{equation}
	\tilde{\cal H}^{\oplus r}:={\rm Span}(\tilde{\cal B}^+).
\end{equation}
Altogether, the physical space ${\cal R}={\cal D}^{\otimes (r+1)}$ is the space of all physical states of the register, while the operational space $\tilde {\cal H}^{\oplus r}\subset{\cal R}$ is the space of those register states that represent valid codewords, and it is isomorphic to the abstract message space ${\cal H}^{\oplus r}$.

A general message is represented by the vector
\begin{equation}\label{nonempty}
	| x\rangle=\sum_{n=0}^r\sum_{i=0}^{k^n-1} x_{n,i}\,|e_i^n\rangle
\end{equation}
with $\sum_{n=0}^r\sum_{i=0}^{k^n-1} |x_{n,i}|^2=1$.
The length operator introduced in section~\ref{sec:lengthop} is here of the form
\begin{equation}\label{lengthopr}
	\hat L:=\sum_{n=0}^{r}n\,\Pi_n,
\end{equation}
because there are at most $r$ digits to constitute a message.
Now we need to know how the projectors $\Pi_n$ are constructed in the operational space $\tilde{\cal H}^{\oplus r}$. For a register state containing a message of sharply defined length, the length eigenvalue $n$ is given by the \emph{number of significant digits} in that register,
\begin{equation}
	\hat L\,|e_i^n\rangle:=n\,|e_i^n\rangle,
\end{equation}
for $0\leq i \leq k^n-1$.
Each projector is then defined by
\begin{equation}
	\Pi_n:=\sum_{i=0}^{k^n-1}|e_i^n\rangle\langle e_i^n|
\end{equation}
and projects onto the space ${\cal H}^{\otimes n}\subset{\cal R}$. 
Note that the \emph{physical length} of each message is always given by the fixed size $(r+1)$ of the register. Only the \emph{significant length} of a message, i.e. the number of digits that constitute a message contained in the register, is in general not sharply defined.
Note further that the particular form of the length operator depends on the realization of the message space. 

In the limit of large $r$ we have $\displaystyle \lim_{r\rightarrow\infty}{\cal H}^{\oplus r}={\cal H}^{\oplus}$, but that space can no longer be embedded into a physical space $\displaystyle{\cal R}={\cal D}^{\otimes\infty}:=\lim_{n\rightarrow\infty}{\cal D}^{\otimes n}$, since the latter is no separable Hilbert space anymore. However, we can think of $r$ as \emph{very large}, such that working in ${\cal H}^{\oplus}$ just means working with a quantum computer having enough memory.

\subsection{Realizing more message spaces}

A code is a map $c:{\cal V}\rightarrow{\cal H}^{\oplus}$ from source states in ${\cal V}$ to codewords in ${\cal H}^{\oplus}$. 
The space ${\cal C}=c({\cal V})$ of all codewords is the \emph{code space} and as a subspace of the general message space ${\cal H}^{\oplus}$ it is just a special message space.
In order to implement a particular code $c$, it is in practice sufficient to realize only the corresponding code space ${\cal C}$ by a physical system. 
Let us realize some important code spaces now. However, we will not discuss the very important class of \emph{error-correcting} code spaces here, since this would go beyond the scope of this paper.

\subsubsection{Block spaces}

An important message space is the \emph{block space} ${\cal H}^{\otimes n}$, that contains messages of fixed length $n$. Block spaces are \emph{the} message spaces of standard quantum information theory. They can directly be realized by a register ${\cal R}={\cal H}^{\otimes n}$ of $n$ digits, e.g. $n$ two-level systems representing one qbit each.

\subsubsection{Prefix spaces}\label{prefix}

Another interesting message space is the space of \emph{prefix codewords} of maximal length $r$. Such a space contains only superpositions of prefix codewords. A set of codewords is \emph{prefix} (or \emph{prefix-free}), if no codeword is the prefix of another codeword.
For example, the set $P_3=\{0,10,110,111\}$ is a set of binary prefix codewords of maximal length 3.
Prefix codewords have one significant advantage:
\begin{itemize}
\item
Prefix codewords are \emph{instantaneous}, that is, sequences of prefix codewords do not need a word separator. The separator can be added while reading the sequence from left to right. A sequence from $P_3$ can be separated like
\begin{equation}
	110111010110 \mapsto 110,111,0,10,110.
\end{equation}
\end{itemize} 
However, there is also a drawback:
\begin{itemize}
\item
Prefix codewords are in general not as short as possible.
\end{itemize}
This is a consequence of the fact that there are in general less prefix codewords than possible codewords. For example, if you want to encode 4 different objects, you can use the prefix set $P_3$ above with maximal length 3. If you renounce the prefix property you can use the set $\{0,1,01,10\}$ with maximal length 2. 

A \emph{prefix space} ${\cal P}_r$ of maximal length $r$ is given by the linear span of prefix codewords of maximal length $r$. For the set $P_3$, the corresponding prefix space is ${\cal P}_3={\rm Span}\{|0\rangle,|10\rangle,|110\rangle,|111\rangle\}$.
The prefix space ${\cal P}_r\subset{\cal H}^{\oplus r}$ can physically be realized by a subspace $\tilde{\cal P}_r$ of the register space ${\cal R}={\cal D}^{\otimes r}$ spanned by the prefix codewords which have been extended by zeroes at the end to fit them into the register. For example, $\tilde{\cal P}_3={\rm Span}\{|000\rangle,|100\rangle,|110\rangle,|111\rangle\}\subset{\cal D}^{\otimes 3}$ is a physical realization of the prefix space ${\cal P}_3$. The length operator measures the \emph{significant length} of the codewords, given by the length of the corresponding prefix codewords.

\emph{Schumacher and Westmoreland} \cite{Schumacher00} as well as \emph{Braunstein et al.} \cite{Braunstein98} used prefix spaces for their implementation of variable-length quantum coding.
However, we will show later on that the significant advantage of prefix codewords in fact vanishes in the quantum case, whereas the disadvantage remains. 

\subsubsection{Neutral-prefix space}\label{neutralpref}

\begin{figure}\vspace{-2em}
	\[{\includegraphics[width=0.3\textwidth]{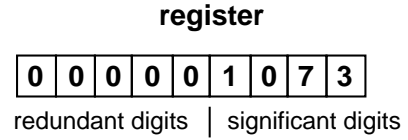}}\]
	\vspace*{-0.5cm}\caption{\small Realizing variable-length messages by neutral-prefix codewords.}\label{neutral}
\end{figure}

A specific code space will be of interest, namely the space of \emph{neutral-prefix codewords}, which we define as follows. 
The $k$-ary representation of a natural number $i$ is denoted by $ Z_k(i)$ (see section~\ref{realizing}). The empty message $\o$ is represented by $ Z_k(0)=\o$. Define an orthonormal basis
\begin{equation}
	{\cal B}_r:=\{| Z_k(0)\rangle,\ldots,| Z_k(k^r-1)\rangle\}
\end{equation}
of variable-length messages of maximal length $r$. 
The length of each basis message $| Z_k(i)\rangle$ is given by
\begin{equation}
	| Z_k(i)|=\lceil\log_k(i+1)\rceil,
\end{equation}
where $\lceil x\rceil$ denotes the smallest integer $\geq x$.
These basis messages span the \emph{$r$-bounded neutral-prefix space}
\begin{equation}
	{\cal N}_r:={\rm Span}({\cal B}_r).
\end{equation}
Note that ${\cal N}_r$ is not equal to the $r$-bounded message space ${\cal H}^{\oplus r}$ as you can see by comparing the dimension 
$\dim{\cal N}_r=k^r$ with $\dim{\cal H}^{\oplus r}=\frac{k^{r+1}-1}{k-1}$.
${\cal N}_r$ is smaller than ${\cal H}^{\oplus r}$, because not all messages of ${\cal H}^{\oplus r}$ are contained in ${\cal N}_r$. For example, the message $|01\rangle$ is in ${\cal H}^{\oplus r}$ but not in ${\cal N}_r$, hence we have
\begin{equation}
	{\cal N}_r\subset{\cal H}^{\oplus r}.
\end{equation}
Now we want to find a physical realization of ${\cal N}_r$. This turns out to be quite easy (see Fig.~\ref{neutral}).
As already noted in section~\ref{realizing}, the $k$-ary representation $ Z_k(i)$ of any natural number $i$ can be extended by leading zeroes to the $r$-extended $k$-ary representation
$ Z_k^r(i):=0\cdots 0 Z_k(i)$.
Take a register ${\cal R}={\cal D}^{\otimes r}$ of $r$ digits with ${\cal D}={\mathbbm C}^k$. Then the set
\begin{equation}
	{\cal B}_{\cal R}:=\{| Z_k^r(0)\rangle,\ldots,| Z_k^r(k^r-1)\rangle\}
\end{equation}
is an orthonormal basis for the register space ${\cal R}$. At the same time it can be regarded as an orthonormal basis for the operational space $\tilde{\cal N}_r$ representing the neutral-prefix space ${\cal N}_r$.
While the \emph{physical length} of each codeword is constantly $r$, the \emph{significant length} is measured by the length operator 
\begin{equation}
	\hat L:=\sum_{n=0}^r n\,\Pi_n,
\end{equation}
with mutually orthogonal projectors
\begin{equation}
	\Pi_n := \sum_{i:\ | Z_k(i)|=n}| Z_k^r(i)\rangle\langle Z_k^r(i)|.
\end{equation}
Note that the so-defined length operator looks different from the one defined in section~\ref{realizing}. While $\hat L$ is always of the same form~(\ref{lengthopr}),  the projectors $\Pi_n$ are different because the operational spaces are different.

The empty message can be defined by 
\begin{equation}
	|\o\rangle:=| Z_k^r(0)\rangle=|0\cdots 0\rangle.
\end{equation}
A general message in $\tilde{\cal N}_r$ is given by
\begin{equation}\label{nr_nonempty}
	| x\rangle =\sum_{i=0}^{k^r-1} x_i\,| Z_k^r(i)\rangle.
\end{equation}
We have realized the neutral-prefix space ${\cal N}_r$ by exhausting the entire register space ${\cal R}$, so the quantum resources are optimally used. In other words: 
\begin{itemize}\item
	All messages in ${\cal N}_r$ are as short as possible.
\end{itemize}
Remember that the physical realization of ${\cal H}^{\oplus r}$ requires one additional digit to represent the beginning or the end of a message. This digit does not contain any message information, it is sort of wasted. For quantum coding, the additional digit may really count, since it would have to be added each time a codeword is stored or transmitted!
Also the prefix space considered in section~\ref{prefix} contains messages which are not as short as possible. You can encode a space ${\cal V}$ of dimension $\dim{\cal V}=4$ by a prefix space spanned by $\{|000\rangle,|100\rangle,|110\rangle,|111\rangle\}$ with corresponding lengths $\{1,2,3,3\}$, but then you need a register of 3 qbits. In contrast to that, ${\cal V}$ can be encoded by a neutral-prefix space spanned by the basis $\{|00\rangle,|01\rangle,|10\rangle,|11\rangle\}$ with corresponding lengths $\{0,1,2,2\}$, and you need a register of only 2 qbits. 
In the operational space $\tilde{\cal N}_r$, the basis messages reveal their length information by simply discarding leading zeroes. That way, not all variable-length messages can be realized, but we save 1 register digit, so ${\cal N}_r$ is a good candidate for variable-length quantum coding.

\section{Data compression}

\subsection{Classical data compression}

Intuitively, compression is achieved when the effort to store or communicate the codewords is minimized. But how can we precisely define that ``effort''? The key idea is the concept of a \emph{raw code}. One can always construct a code for $\Omega$ by inventing a new letter for each single object. Such a classical \emph{raw code} is a code $c:\Omega\rightarrow{\cal A}$ for some alphabet ${\cal A}$ of the same size as $\Omega$. 
The chinese writing is a fairly good illustration of a raw code. There are up to 50,000 letters representing a manifold of abstract and concrete things, e.g. the ``noise of a running horse''.
The length of the code is minimized to 1, but the encoding and decoding machines will need a large memory to remember all the letters. Obviously, a raw code does not compress at all, so it is a good idea to set the effort of communication in relation to the \emph{raw information content} of $\Omega$ (similiar notion in \cite{Preskill} p.71, and interestingly similiar also to the Boltzman entropy of a microcanonical ensemble), defined by
\begin{equation}\label{raw_inf}
	I_0(\Omega):=\log_2|\Omega|.
\end{equation}
$I_0(\Omega)$ represents the number of \emph{binary digits (bits)} needed to enumerate the elements of $\Omega$.
This motivates the following definition. The \emph{code information content} of an individual object in an arbitrary set $\Omega$ for a given $k$-ary code $c:\Omega\rightarrow{\cal A}^+$ is defined as
\begin{equation}\label{code_inf}
	I_c(x):=\log_2 k\cdot L_c(x),\quad x\in\Omega,
\end{equation}
where $L_c(x)$ denotes the length of the codeword $c(x)\in{\cal A}^+$.
$I_c(x)$ represents the number of bits needed to describe the object $x$ by the code $c$.
For a raw code $c:\Omega\rightarrow{\cal A}$, definition~(\ref{code_inf}) gives the raw information content for every object $x\in\Omega$. 
A few remarks about the code information:

1) The code information is defined for \emph{things}, not for \emph{strings}. Of course, things may sometimes also be strings. If so, one can define the \emph{direct information} of a string $x^n$ over an alphabet ${\cal A}$ as
\begin{equation}\label{direct_inf}
	I(x^n):=n\,\log_2|{\cal A}|.
\end{equation}

2) The code information $I_c$ is \emph{code dependent}, reflecting the philosophy that there is no information contained in an object without a code giving it some \emph{meaning}. The codeword "XWF\$\%\&\$ FggHz((" may be a random sequence of letters or may in a certain code represent the first digits of $\pi$ or in another code the beginning of a Mozart symphony.

Now let there be a probability distribution $p$ on $\Omega$. 
We can define the code information of the ensemble $\Sigma=\{p,\Omega\}$ as the average of~(\ref{code_inf}),
\begin{equation}\label{ens_inf}
	I_c(\Sigma):=\log_2 k\,\sum_{ x\in\Omega}p( x)\,L_c( x).
\end{equation}
\emph{Compression} means reducing the code information of the ensemble. We can define the \emph{compression rate} achieved by a code $c$ on the ensemble $\Sigma$ by
\begin{equation}\label{ens_comp}
	{\rm R}_c(\Sigma):=\frac{I_c(\Sigma)}{I_0(\Omega)},
\end{equation}
A code $c:\Omega\rightarrow{\cal C}$ is \emph{compressive} on $\Sigma$ if and only if
\begin{equation}\label{ens_comp_test}
	{\rm R}_c(\Sigma)<1\quad\text{i.e.}\quad I_c(\Sigma)<I_0(\Omega).
\end{equation}

\subsection{Quantum data compression}

Now that we have a classical definition of compression, the next step is to translate these concepts to the quantum case.
Again, the key is the raw information, i.e. the size of a non-compressed message, so let us look for its quantum analogue. The raw information~(\ref{raw_inf}) of a set $\Omega$ is $I_0(\Omega)=\log_2|\Omega|$ because we need $|\Omega|$ distinct letters to encode each element of $\Omega$ by a raw code. Interpreting $\Omega$ as an orthonormal basis for a Hilbert space ${\cal V}$, the raw information of ${\cal V}$ is also $\log_2|\Omega|$, because we still need $|\Omega|$ distinguishable letters to represent each element of the space ${\cal V}$. Since $|\Omega|=\dim{\cal V}$, we define the \emph{quantum raw information} of a space ${\cal V}$ as 
\begin{equation}\label{qraw_inf}
	I_0({\cal V}):=\log_2(\dim {\cal V}).
\end{equation}
So the quantum raw information $I_0$ corresponding to a space ${\cal V}$ equals the fixed number of qbits needed to represent all states in ${\cal V}$.

Now, for a given $k$-ary code $c:{\cal V}\rightarrow{\cal H}^{\oplus}$ represented by an encoder $C$, the \emph{code information operator} can be defined as
\begin{equation}\label{inf_op}
	\hat I_c:=\log_2 k\cdot\hat L_c,
\end{equation}
where $\hat L_c:=C^{-1}\,\hat L\,C$ is the length operator measuring the length of the codeword for a source vector in ${\cal V}$. If the code is based on a qbit alphabet, $\hat I_c$ measures the number of qbits forming the code message, hence the measuring unit of $\hat I_c$ is ``1 qbit''.
In analogy to~(\ref{direct_inf}), we define the \emph{direct information operator} acting on the message space ${\cal H}^{\oplus}$ by
\begin{equation}\label{direct_inf_op}
	\hat I:=\log_2 k\cdot\hat L.
\end{equation}
In short, the code information operator is defined in an arbitrary Hilbert space ${\cal V}$ and depends on a quantum code $c:{\cal V}\rightarrow{\cal H}^{\oplus}$, while the direct information operator is defined in a message space ${\cal H}^{\oplus}$ without referring to a quantum code. For a given code, the relation between both operators is
\begin{equation}
	\hat I_c = C^{-1}\,\hat I\,C.
\end{equation}
Now you want to compress a codeword by removing redundant quantum digits. The number of quantum digits carrying information is given by the base length of the codeword. All other digits are redundant and can be removed without loss of information.
This motivates the definition of the \emph{code information} of a state $|x\rangle\in{\cal V}$ respecting a code $c$ by
\begin{equation}\label{qinf}
	\underline I_c(x):=\log_2 k\cdot\underline L_c(x),
\end{equation}
where $\underline L_c(x)=\underline L(c(x))$ is the base length of the codeword for $|x\rangle$. $\underline I_c(x)$ represents the number of qbits needed to describe the state $|x\rangle$ by the code $c$. This value must be distinguished from the \emph{expected} number of qbits $I_c(x)=\langle x|\hat I_c|x\rangle$ that is found by performing a length measurement on the codeword for $|x\rangle$. In the classical case, the difference vanishes.

Now suppose you want to encode an ensemble $\Sigma=\{p,{\cal X}\}$ of states $| x\rangle\in {\cal X}$ that span the source space ${\cal V}$. Each individual message $| x\rangle$ can be compressed to $\underline  I_c( x)$ qbits, so the entire ensemble $\Sigma$ will on the average be compressed to the code information
\begin{equation}\label{qens_inf}
	 \underline I_c(\Sigma)
	:=\log_2 k\,\sum_{ x\in{\cal X}}p( x)\,\underline L_c( x).
\end{equation}
The compression rate can then be defined by
\begin{equation}\label{qens_comp}
	{\rm R}_c(\Sigma):=\frac{\underline I_c(\Sigma)}{I_0({\cal V})}.
\end{equation}
A code $c$ is compressive on the ensemble $\Sigma$, if and only if
\begin{equation}\label{qens_comp_test}
	{\rm R}_c(\Sigma)<1\quad\text{i.e.}\quad
	\underline I_c(\Sigma)<I_0({\cal V}).
\end{equation}
Note that these definitions only apply to \emph{lossless} codes. The lossy case is not considered here.

\section{No-go theorems}

Of course, \emph{lossy} compression is always possible. But let us look for some statements about \emph{lossless codes}.
The first three of the following no-go theorems are also known in classical information theory and are easily transferred to the quantum case by general reasoning. However, we show them by applying the tools developped in this paper.
The last theorem is genuinely quantum with no classical analogue.

\subsection{No lossless compression by block codes}\label{noblock}

A code is a \emph{block code} if all codewords have the same length, else it is a \emph{variable-length code}.
Unfortunately, lossless block codes do not compress.
Take an arbitrary ensemble $\Sigma=\{p,{\cal X}\}$ with ${\cal X}\subset{\cal V}$ and any lossless $k$-ary block code $c:{\cal V}\rightarrow{\cal H}^{\otimes n}$.
Let ${\cal B}_{\cal V}$ and ${\cal B}_n$ be orthonormal basis sets of ${\cal V}$ and ${\cal H}^{\otimes n}$, respectively. 
In order to find for every basis vector $|\omega\rangle\in{\cal B}_{\cal V}$ a code basis vector $|c(\omega)\rangle\in{\cal B}_n$, the code must fulfill $\dim{\cal V}\leq \dim{\cal H}^{\otimes n}=k^n$. For every $| x\rangle\in{\cal X}$, the corresponding codeword $|c( x)\rangle$ has sharp length $L(x)=n$, hence
\begin{eqnarray}
	\underline I_c(\Sigma)&=&\log_2 k\,\sum_{ x\in{\cal X}}p( x)\,
	 L_c( x)=\log_2 k\cdot n=\log_2(k^n)\\
	&\geq&\log_2(\dim{\cal V})=I_0({\cal V}),
\end{eqnarray}
which violates condition~(\ref{qens_comp_test}).
This implies that
there is no lossless compressing block code.
By choosing mutually orthogonal source states one can derive the analogue statement for the classical case.

For \emph{long strings} emitted by a memoryless source, block codes can achieve \emph{almost lossless} compression by encoding only \emph{typical subspaces}. The quantum code performing this type of lossy compression is known as the \emph{Schumacher code} \cite{Schumacher96}.
The only way to compress messages without loss of information is by use of a variable-length code. 
In order to achieve compression, more frequent objects must be encoded by shorter messages, less frequent objects by longer messages, so that the average length of the codes is minimized.
This is the general rule of lossless data compression.

\subsection{No lossless compression by changing the alphabet}

Trying to achieve compression by using a different alphabet does not work. 

A code $c:{\cal H}_A^{\otimes n}\rightarrow{\cal H}_B^{\otimes m}$ that transforms messages over some letter space ${\cal H}_A$ into messages over some letter space ${\cal H}_B$ is lossless only if $\dim{\cal H}_A^{\otimes n}\leq\dim{\cal H}_B^{\otimes m}$, which implies that
\begin{eqnarray}
	I_0({\cal V})&=&n\,\log_2(\dim{\cal H}_A)\\
	&\leq& m\,\log_2(\dim{\cal H}_B)=\underline I_c(x),
\end{eqnarray}
for every $|x\rangle\in{\cal H}_A$. So for every ensemble $\Sigma=\{p,{\cal X}\}$ of messages $|x\rangle\in{\cal H}_A^m$, we have $\underline I_c(\Sigma)=\underline I_c(x)\geq I_0({\cal V})$, which violates condition~(\ref{qens_comp_test}).
By choosing mutually orthogonal source states, one can derive the analogue statement for the classical case.
This paper \emph{looks} probably much shorter when written in chinese symbols. However, the effort of communication that is expressed by the code information $I_c$, would not be reduced.

\subsection{No universal lossless compression}

We have seen that it is not possible to compress messages without loss of information by using a block code or by using a different letter space. Now we will see that no code can compress \emph{all} messages without loss of information. 

Say you have a space ${\cal H}^{\otimes n}$ of \emph{block messages} of fixed length $r$ and you want to compress all of them by use of a variable-length code $c:{\cal H}^{\otimes r}\rightarrow{\cal H}^{\oplus s}$ with $s<r$. The code can only be lossless if
\begin{equation}
	\dim{\cal H}^{\otimes r}\leq\dim{\cal H}^{\oplus s}.
\end{equation}
But since $\dim{\cal H}^{\otimes r}=k^r$ and $\dim{\cal H}^{\oplus s}=\frac{k^{s+1}-1}{k-1}$, we have
\begin{eqnarray}
	k^r&\leq&\frac{k^{s+1}-1}{k-1}\\
	\Rightarrow\quad k^{r+1}&\leq& k^{s+1}+k-1
\end{eqnarray}
which is wrong for $r\geq s$ and $k>1$, so you cannot compress all block messages of a given length.
Now say you have a space ${\cal H}^{\oplus r}$ of \emph{variable-length messages} with maximal length $r$. Assume that there is a universal lossless code $c$ that reduces the length of all messages in ${\cal H}^{\oplus r}$. The code can only be lossless if $\dim{\cal H}^{\oplus r}\leq\dim{\cal H}^{\oplus s}$, which is obviously wrong for $r>s$, so you cannot compress all variable-length messages with a given maximal length. Concluding, there is no universal lossless compression that reduces the size of all messages. Some messages are unavoidably lengthened by a lossless code.
By choosing mutually orthogonal source states, one can derive the analogue statement for the classical case.

\subsection{No lossless compression of unknown messages}\label{oracle}

Now we come to a no-compression theorem that is typically quantum. In quantum mechanics there is a profound difference between a \emph{known} and an \emph{unknown} state. For example, a known state can be cloned (by simply preparing another copy of it), whereas an unknown state cannot be cloned. 

Assume that there is a lossless quantum compression algorithm $c:{\cal H}^{\otimes r}\rightarrow{\cal H}^{\oplus s}$ that compresses messages of fixed length $r$ to variable-length messages of maximal length $s$. As we have seen in the last section, a lossless code cannot compress \emph{all} messages, so $s>r$. Now there is an oracle that hands you an arbitrary message $| x\rangle=\sum_{i=1}^n  x_i\,|\omega_i\rangle$ where the $|\omega_i\rangle\in{\cal H}^{\oplus r}$ are mutually orthogonal states. The algorithm encodes the message $|x\rangle$ into
$|c(x)\rangle=\sum_{i=1}^n  x_i\,|c(\omega_i)\rangle$. Even if all the codeword components $|c(\omega_i)\rangle$ have determinate length $L_c(\omega_i)$, the total codeword $|c( x)\rangle$ has in general indeterminate length. If you want to remove redundant digits without loss of information, you must know at least an upper bound for its base length, i.e. the length of its longest component. Since you do not know the source message $|x\rangle$, you do not know the base length of its encoding $|c(x)\rangle$, so you have to assume the maximal length $s$. Since $s>r$, no compression is achieved. 
The same argument applies to quantum compression algorithms $c:{\cal H}^{\oplus r}\rightarrow{\cal H}^{\oplus s}$ compressing variable-length messages of maximal length $r$ to variable-length messages of maximal length $s$.

We conclude that lossless compression of unknown quantum messages is in general impossible.
This statement is not true for the classical case. A classical message is not disturbed by a length measurement, so it \emph{can} in principle be compressed without loss of information.
It would have been nice to compress a quantum hard disk without loss of information just like a classical hard disk, but this cannot be accomplished in general.

Now that we have found a lot of impossible things to do with quantum messages, it is time to look for the possible things.

\section{Lossless compressing codes}

The intention of using compressing codes is to minimize the effort of communication between two parties: one who \emph{prepares, encodes, compresses and sends} the messages and one who \emph{receives, decompresses, decodes} and possibly \emph{reads} them. So it's time for Alice and Bob to enter the scene. Alice is preparing source messages $|x\rangle\in{\cal V}$ and encodes them into codewords $|c(x)\rangle\in{\cal H}^{\oplus r}$ by applying the encoder $C$. She compresses the codewords by removing redundant quantum digits and sends the result to Bob, who receives them and decompresses them by appending quantum digits. After that he can decode the messages by applying the decoder $D$ and read them or use them as an input for further computations. The communication has been lossless, if the decoded message equals the source message. Note that it is not required for Bob to \emph{read} the message he received! In fact, if Bob wants to use the message as an input for a quantum computer, he even \emph{must not} do that, else he will potentially lose information.
We require Alice to \emph{know} which source messages she prepares, otherwise no lossless compression is possible, as we have seen in the previous section.

\subsection{Why prefix quantum codes are not very useful}

In classical information theory, prefix codes are favored for lossless coding. The reason is that they are \emph{instantaneous}, which means that they carry their own length information (see section~\ref{prefix}). Prefix codewords can be sent or stored without a separating signal between them. The decoder can add word separators (``commas'') while reading the sequence from left to right.
Whenever a string of letters yields a valid codeword, the decoder can add a 
comma and proceed. After all, a continuous stream of letters is separated into valid codewords.

Prefix codewords can be separated while \emph{reading} the sequence, but in the quantum case this is potentially a very bad thing to do. Reading a stream of quantum letters means in general \emph{disturbing} the message all the time. Therefore, the length information is generally not available. Furthermore, prefix codewords are in general \emph{longer} than non-prefix codewords, because there are less prefix codewords of a given maximal length than possible codewords. Hence, by using prefix codewords qbits are wasted to encode length information which is unavailable anyway. We conclude that prefix quantum codes are practically not very useful.

\subsection{A classical side-channel}\label{side}

One could try to encode length information in a different quantum channel, as proposed by \emph{Braunstein et al.} \cite{Braunstein98} (unnecessarily they used prefix codewords anyhow). But that does not fix the problem. Whatever one does, reading out length information about different components of a variable-length codeword equals a length measurement and hence means disturbing the message.
Though there should be \emph{some} way to make sure where the codewords have to be separated, else the message cannot be decoded at all. Here is an idea: Use a \emph{classical side-channel} to inform the receiver where the codewords have to be separated. This has two significant advantages:
\begin{itemize}
\item
If the length information equals the base length of the codeword, the message is not disturbed and can be losslessly transmitted and decoded.
\item
Abandoning the prefix condition, shorter codewords can be chosen, such that the quantum channel is used with higher efficiency.
\end{itemize}
\begin{figure}\vspace{-2em}
	\[{\includegraphics[width=0.4\textwidth]{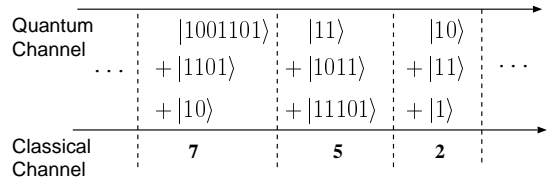}}\]
	\vspace*{-0.5cm}\caption{\small Storing length information in a classical side-channel.}\label{lengths}
\end{figure}
Let us give an example (see Fig.~\ref{lengths}). 
Alice wants to send a message $|x_1\rangle$ which is encoded into the codeword
$|c(x_1)\rangle=\frac{1}{\sqrt{3}}(|1001101\rangle+|1101\rangle+|10\rangle)$. The base length of $|c(x_1)\rangle$ is $7$, so she submits that information through the classical channel. Dependent on which realization of variable-length messages Alice and Bob have agreed to use, Alice sends enough qbits (at least 7) representing the codeword $|c(x_1)\rangle$ through the quantum channel. The next codeword is 
$|c(x_2)\rangle=\frac{1}{\sqrt{3}}(|11\rangle+|1011\rangle+|11101\rangle)$. The base length of $|c(x_2)\rangle$ is $5$, so Alice sends the length information ``5'' through the classical channel and enough qbits (at least 5) representing the codeword $|c(x_2)\rangle$ through the quantum channel. She proceeds like that with all following messages. On Bob's side, there is a continuous stream of qbits coming through the quantum channel and a continuous stream of classical bits coming through the classical channel. Bob can read out the classical length information, separate the qbits into the specified blocks and apply the decoder to each codeword. After all, Bob obtains all source messages without loss of information.

\subsection{How much compression?}

\subsubsection{Lower bound}\label{lower}

How much compression can maximally be achieved by using the method sketched in section~\ref{side}?
Say Alice has an ensemble $\Sigma=\{p,{\cal X}\}$ of $m=|{\cal X}|$ messages $|x_i\rangle\in{\cal X}$, $i=1,\ldots,m$ that she wants to encode by $k$-ary codewords. The source space ${\cal V}$ is spanned by the elements of ${\cal X}$, thus ${\cal V}:={\rm Span}({\cal X})$, and has dimension $d:=\dim{\cal V}$. Alice fixes a basis set ${\cal B}_{\cal V}$ of $d$ orthonormal vectors $|\omega_i\rangle$, $i=1,\ldots,d$.
The ensemble $\Sigma$ corresponds to the message matrix
\begin{equation}
	\sigma :=\sum_{i=1}^m p(x_i)\,|x_i\rangle\langle x_i|
		=\sum_{i,j=1}^d \sigma_{ij}\,|\omega_i\rangle\langle \omega_j|,
\end{equation}
with $\sigma_{ij}:=\langle \omega_i|\sigma|\omega_j\rangle$ and $\sum_{i=1}^d\sigma_{ii}=1$.
The source messages are encoded by the isometric map $c:{\cal V}\rightarrow{\cal H}^{\oplus}$, defined by
\begin{equation}
	|\omega_i\rangle\stackrel{c}{\longmapsto} |c(\omega_i)\rangle,\quad i=1,\ldots d.
\end{equation}
The code space is $k$-ary, which means that $k=\dim{\cal H}$.
Let each codeword $|c(\omega_i)\rangle$ have determinate length $L_c(\omega_i)$, such that the code length operator $\hat L_c$ on ${\cal V}$ is orthogonal in the basis ${\cal B}_{\cal V}$ and reads
\begin{equation}
	\hat L_c =\sum_{i=1}^d L_c(\omega_i)\,|\omega_i\rangle\langle\omega_i|.
\end{equation}
The codewords $|c(\omega_i)\rangle$ are not necessarily prefix, because Alice can encode the length information about each codeword in a classical side-channel. In order for the transmission to be lossless, she has to transmit the base length $\underline L_c(x_i)$ of each codeword corresponding to the source message $|x_i\rangle$. The base length is at least as long as the expected code length of the codeword, hence
\begin{equation}
	\underline L_c(x_i)\geq\langle x_i|\hat L_c|x_i\rangle.
\end{equation}
Now we are interested in the average base length, since this determines the compression rate. The average base length is bounded from below by
\begin{eqnarray}
	\underline L_c(\Sigma)&=&\sum_{i=1}^m p(x_i)\,\underline L_c(x_i)\\
	&\geq& \sum_{i=1}^m p(x_i)\,\langle x_i|\hat L_c|x_i\rangle
	={\rm Tr}\{\sigma\,\hat L_c\}\\
	&=&\sum_{i=1}^m \sigma_{ii}\,L_c(\omega_i)\label{av_base}.
\end{eqnarray}
Now we perform the following trick. 
As already stated, non-prefix codewords can be chosen shorter than (or at most as long as) prefix codewords.
Consider an arbitrary prefix code $c'$, then
\begin{equation}
	L_{c'}(\omega_i)=L_c(\omega_i)+l_{c'}(\omega_i)\geq L_c(\omega_i),
\end{equation}
where $l_{c'}(\omega_i)\geq 0$ is the length difference between the prefix and the non-prefix codeword for $|\omega_i\rangle$.
Prefix codes, just like all uniquely decodable symbol codes, have to fulfill the \emph{Kraft inequality} \cite{CoverThomas91,MacKay}
\begin{equation}
	\sum_{i=1}^d k^{-L_{c'}(\omega_i)}\leq 1.
\end{equation}
Since the code length operator $\hat L_{c'}$ is orthogonal in the basis ${\cal B}_{\cal V}$, we can express the above condition by the \emph{quantum Kraft inequality}
\begin{equation}
	{\rm Tr}_{\cal V}\{k^{-\hat L_{c'}}\}\leq 1,
\end{equation}
where $\hat L_{c'}:=\hat L_c+\hat l_{c'}$ and
\begin{equation}
	\hat l_{c'}:=\sum_{i=1}^d l_{c'}(\omega_i)\,|\omega_i\rangle\langle\omega_i|.
\end{equation}
The quantum Kraft inequality was derived for the first time by \emph{Schumacher and Westmoreland} \cite{Schumacher00}. 
Here, the quantum Kraft inequality requires that
\begin{equation}
	Q:=\sum_{i=1}^d k^{-L_c(\omega_i)-l_{c'}(\omega_i)}\leq1.
\end{equation}
Now define \emph{implicit probabilities}
\begin{equation}
	q(\omega_i):=\frac1Q\,k^{-L_c(\omega_i)-l_{c'}(\omega_i)},
\end{equation}
which can be rewritten as
\begin{equation}
	L_c(\omega_i)=-\log_k q(\omega_i)-\log_k Q-l'(\omega_i).
\end{equation}
Summing over the $\sigma_{ii}$ yields
\begin{equation}
	\sum_{i=1}^d\sigma_{ii}\,L_c(\omega_i)
	=-\sum_{i=1}^d\sigma_{ii}\log_k q(\omega_i)-\log_k Q-l',
\end{equation}
where
\begin{equation}
	l':=\sum_{i=1}^d \sigma_{ii}\,l_{c'}(\omega_i)
	={\rm Tr}\{\sigma\,\hat l_{c'}\}
\end{equation}
is the average additional length.
The inequality~(\ref{av_base}) can now be expressed by
\begin{eqnarray}
	\underline L_c(\Sigma)&\geq& -\sum_{i=1}^d \sigma_{ii}\,
	\log_k q(\omega_i)-\log_k Q-l'
\end{eqnarray}
Gibbs' inequality implies that
\begin{equation}\label{gibbs}
	\underline L_c(\Sigma)\geq -\sum_{i=1}^d \sigma_{ii}\,\log_k \sigma_{ii}-\log_k Q-l'.
\end{equation} 
The von-Neumann entropy of the message matrix $\sigma$ cannot  decrease by a non-selective projective measurement in the basis ${\cal B}_{\cal V}$, hence
\begin{eqnarray}\label{nonincrease}
	S(\sigma)\leq S(\sigma'),
\end{eqnarray}
where
\begin{eqnarray}
	\sigma'&:=&\sum_{i=1}^d |\omega_i\rangle\langle\omega_i|
		\sigma|\omega_i\rangle\langle\omega_i|
		=\sum_{i=1}^d \sigma_{ii}|\omega_i\rangle\langle\omega_i|.
\end{eqnarray}
Since
\begin{eqnarray}
	S(\sigma')=-\sum_{i=1}^d \sigma_{ii}\,\log_2\sigma_{ii}
	=-\log_2 k\,\sum_{i=1}^d \sigma_{ii}\,\log_k\sigma_{ii},
\end{eqnarray}
relation~(\ref{nonincrease}) states that
\begin{equation}\label{vn}
	-\sum_{i=1}^d \sigma_{ii}\,\log_k \sigma_{ii}\geq\frac1{\log_2 k}\,S(\sigma).
\end{equation}
Using~(\ref{vn}) together with the Kraft inequality $Q\leq 1$, relation~(\ref{gibbs}) transforms into
\begin{equation}
	\log_2 k\cdot\big\{\underline L_c(\Sigma)+l'\big\}
	\geq S(\sigma)-\log_k Q\geq S(\sigma).
\end{equation} 
Recalling the definition of the code information~(\ref{qens_inf})
and defining the length information that can be drawn into the classical side-channel by
\begin{equation}
	I':=\log_2 k\cdot l',
\end{equation}
we finally arrive at the lower bound relation
\begin{equation}\label{lower_bound}
	\underline I_c(\Sigma)+I'\geq S(\sigma).
\end{equation}
If $c$ is a uniquely decodable symbol code, e.g. a prefix code, we have $I'=0$.
Inequality~(\ref{lower_bound}) states that the ensemble $\Sigma$ can be losslessly compressed not below $S(\sigma)$ qbits.
However, by drawing length information into a classical side-channel it is possible to reduce the average number of qbits passing through the quantum channel \emph{below} the von-Neumann entropy. We will give an example later on where this really happens.

\subsubsection{Upper bound}

Let us look for an upper bound for the compression that can be achieved. 
In order to encode every source vector in ${\cal V}$ by a $k$-ary code, we need at most
\begin{equation}
	\underline L_c(x)\leq\lceil \log_k(\dim{\cal V})\rceil
	\leq\log_k(\dim{\cal V})+1
\end{equation}
digits. Using $\log_a x=\log_a b\cdot\log_b x$, we have
\begin{equation}\label{rawbound}
	\underline I_c(\Sigma)\leq\log_2(\dim{\cal V})+\log_2 k.
\end{equation}
This upper bound is neither very tight nor is it related to the von-Neumann entropy. However, our efforts to find a more interesting upper bound were not successful. It remains an open question to find such a bound and hence a quantum mechanical generalization to Shannon's theorem \cite{Shannon48},
\begin{equation}\label{shannonlossless}
	H(\Sigma)\leq I_c(\Sigma)\leq H(\Sigma)+\log_2 k,
\end{equation}
which looks more familiar for $k=2$, such that $\log_2 k=1$ and $I_c(\Sigma)=L_c(\Sigma)$.

\subsection{Quantum Morse codes}\label{qmorse}

One way to avoid a classical side-channel is to leave a \emph{pause} between the quantum codewords, which equals an additional orthogonal ``comma state''. Such a code is a quantum analogue to the \emph{Morse code}, where the codewords are also separated by a pause, in order to avoid prefix codewords. Of course, the codewords \emph{plus} the pause are prefix. Due to the close analogy one could speak of \emph{quantum Morse codes}.
Here, the information $I'$ needed for the comma state is independent from the statistics, because the comma state must be sent after each letter codeword, no matter which one. In contrast to that, $I'$ is in general dependent from the statistics. If one transmits the length of each codeword through a classical side-channel, one can use a Huffman code to find shorter codewords for more frequent length values. Such is done in the following compression scheme.

\section{A lossless compression scheme}\label{explicit}

Let us construct an explicit coding scheme that realizes lossless quantum compression.

\subsection{Preparations}

Alice and Bob have a quantum computer on both sides of the channel. They both allocate a register of $r$ $k$-ary quantum digits, whose physical space is given by ${\cal R}={\cal D}^{\otimes r}$ with ${\cal D}={\mathbbm C}^k$.
They agree to use neutral-prefix codewords (see section~\ref{neutralpref}) to implement variable-length coding, hence the message space is ${\cal N}_r$ of dimension $k^r$ and is physically realized by the operational space $\tilde{\cal N}_r={\cal R}$. 
Alice is preparing source messages $|x_i\rangle, i=1,\ldots,m$ from a set ${\cal X}$. The space spanned by these messages is the source space ${\cal V}={\rm Span}({\cal X})$. 
Alice prepares each message $|x\rangle\in{\cal X}$ with probability $p(x)$, which gives the ensemble $\Sigma:=\{p,{\cal X}\}$. 
She encodes the source messages into variable-length codewords $|c(x)\rangle\in{\cal N}_r$ of maximal length $r$. 
If the dimension of ${\cal V}$ is given by $d:=\dim{\cal V}$, then the length of the register must fulfill
\begin{equation}
	r \geq \lceil \log_k d\rceil.
\end{equation}
If the set ${\cal X}$ is linearly dependent, Alice creates a set $\tilde{\cal X}={\cal X}$, removes the most probable message from $\tilde{\cal X}$ and puts it into a list $\boldsymbol M$. Next, she removes again the most probable message from $\tilde{\cal X}$, appends it to the list $\boldsymbol M$ and checks if the list is now linearly dependent. If so, she removes the last element from $\boldsymbol M$ again. Then she proceeds with removing the next probable message from $\tilde{\cal X}$ and appending it to $\boldsymbol M$, checking for linearly dependence, and so on. In the end she obtains a list
\begin{equation}
	\boldsymbol M=(| x_1\rangle,\ldots,| x_d\rangle)
\end{equation}
of linearly independent source messages from ${\cal X}$, ordered by decreasing probability, such that
$p(x_i)\geq p(x_j)$ for $i\leq j$.
She performs a \emph{Gram-Schmidt} orthononormalization on the list $\boldsymbol M$, giving a list $\boldsymbol B$ of orthornormal vectors $|\omega_i\rangle$, defined by
\begin{eqnarray}
	|\omega_1\rangle &:=& | x_1\rangle,\\
	|\omega_i\rangle &:=& N_i\,\Big[{\mathbbm 1}-\sum_{j=1}^{i-1}
		|\omega_j\rangle\langle\omega_j|\Big]|x_i\rangle,
\end{eqnarray}
with $i=2,\ldots,d$ and suitable normalization constants $N_i$.
The elements of $\boldsymbol B$ form an orthonormal basis ${\cal B}_{\cal V}$ for the source space ${\cal V}$.
Now she assigns codewords 
\begin{equation}
	|c(\omega_i)\rangle:=| Z_k^r(i-1)\rangle,\quad i=1,\ldots,d.
\end{equation}
of increasing significant length
\begin{equation}
	L_c(\omega_i)=\lceil \log_k (i)\rceil.
\end{equation}
Note that the first codeword is the empty message $|\o\rangle=| Z_k^r(0)\rangle=|0\cdots 0\rangle$, which does not have to be sent through the quantum channel at all. Instead, nothing is sent through the quantum channel and a signal representing ``length 0'' is sent through the classical channel. 
Alice implements the encoder
\begin{equation}
	C:=\sum_{i=1}^d |c(\omega_i)\rangle\langle\omega_i|,
\end{equation}
by a gate array on ${\cal R}$.
Then she calculates the base lengths of the codewords,
\begin{equation}
	\underline L_c(x)=\max_{i=1,\ldots,d}
	\{L_c(\omega_i)\mid |\langle\omega_i|x\rangle|^2> 0\},
\end{equation}
for every message $|x\rangle\in{\cal X}$ and writes them into a table. 
The classical information is compressed using Huffman coding of the set of distinct base length values ${\cal L}=\{L_c(\omega_1),\ldots,L_c(\omega_d)\}$. Alice constructs the Huffman codeword to each length $l\in{\cal L}$ appearing with probability
\begin{equation}
	p_l=\sum_{x:\,\underline L_c(x)=l}p(x),
\end{equation}
and writes them into a table.
At last, Alice builds a gate array realizing the decoder $D=C^{-1}$ and gives it to Bob. 
For the classical channel she hands the table with the Huffman codewords for the distinct lengths to Bob.
Now everything is prepared and the communication can begin.

\subsection{Communication protocol}

Alice prepares the message $|x\rangle\in{\cal X}$ and applies the encoder $C$ to obtain $|c(x)\rangle$. She looks up the corresponding code base length $\underline L_c(x)$ in the table. If $\underline L_c(x)<r$, she truncates the message to $\underline L_c(x)$ digits by removing $r-\underline L_c(x)$ leading digits.
She sends the $\underline L_c(x)$ digits through the quantum channel and the length information $\underline L_c(x)$ through the classical channel. Then she proceeds with the next message.

For any message $|x\rangle$ Alice sends, Bob receives the length information $\underline L_c(x)$ through the classical channel and $\underline L_c(x)$ digits through the quantum channel. He adds $r-\underline L_c(x)$ quantum digits in the state $|0\rangle$ at the beginning of the received codeword. He then applies the decoder $D$ and obtains the original message $|x\rangle$ with perfect fidelity.
Note that Alice can send \emph{any} message from the source message space ${\cal V}$, the protocol will ensure a lossless communication of the message. For such arbitrary messages, however, compression will in general not be achieved, since the protocol is only adapted to the particular ensemble $\Sigma$.
Also, Bob can as well store all received quantum digits on his quantum hard disk and the received length information on his classical hard disk, and go to bed. The next day, he can scan the classical hard disk for length information and separate and decode the corresponding codewords on the quantum hard disk.
The protocol works as well for online communication as for data storage.

\subsection{An explicit example}

Alice and Bob want to communicate vectors of a 4-dimensional Hilbert space
${\cal V}={\rm Span}\{|0\rangle,|1\rangle,|2\rangle,|3\rangle\}$, where we use the row notation in the following.
Alice decides to use the (linearly dependent) source message set
\begin{equation}
	{\cal X}=\{|a\rangle,|b\rangle,|c\rangle,|d\rangle,|e\rangle,
	|f\rangle,|g\rangle,|h\rangle,|i\rangle,|j\rangle\},
\end{equation}
whose elements are given by
\begin{eqnarray}
	|a\rangle&=&\frac1{2}(1, 1, 1, 1)\\
	|b\rangle&=&\frac1{\sqrt5}(1, 2, 1, 1)\\
	|c\rangle&=&\frac1{\sqrt6}(1, 3, 1, 1)\\
	|d\rangle&=&\frac1{\sqrt7}(1, 4, 1, 1)\\
	|e\rangle&=&\frac1{\sqrt2}(1, 0, 1, 0)\\
	|f\rangle&=&\frac1{\sqrt3}(2, 0, 1, 0)\\
	|g\rangle&=&\frac1{2}(3, 0, 1, 0)\\
	|h\rangle&=&\frac1{\sqrt2}(0, 1, 0, 1)\\
	|i\rangle&=&\frac1{\sqrt3}(0, 2, 0, 1)\\
	|j\rangle&=&\frac1{2}(0, 3, 0, 1)\\
\end{eqnarray}
and which are used with the probabilities
\begin{eqnarray}
	p(a)&=&0.6,\quad p(b)=p(c)=p(d)=0.1,\\
	p(e)&=&\ldots=p(j)=\frac{0.3}{3}.
\end{eqnarray}
The Shannon entropy of the ensemble $\Sigma=\{p,{\cal X}\}$ is
\begin{equation}\label{shannonsigma}
	H(\Sigma)=2.02945,
\end{equation}
and the classical raw information~(\ref{raw_inf}) reads
\begin{equation}
	I_0({\cal X})=\log_2|{\cal X}|=3.32193,
\end{equation}
which gives an optimal classical compression rate of $R=H/I_0=0.610924$.
If Bob knows Alice's list of possible messages, then this rate could in the optimal case be achieved by pure classical communication. However, Bob does not know the list and classical communication is not the task here.
The message matrix $\sigma=\sum_{x\in{\cal X}}p(x)|x\rangle\langle x|$, given by
\begin{equation}
	\sigma=\begin{pmatrix}
	 0.214549 & 0.224624 & 0.197882 & 0.177882 \\ 0.224624 & 0.40302 & 
    0.224624 & 0.244624 \\ 0.197882 & 
    0.224624 & 0.191216 & 0.177882 \\ 0.177882 & 
    0.244624 & 0.177882 & 0.191216 \\   
	\end{pmatrix}
\end{equation}
has von-Neumann entropy
\begin{equation}\label{vonneumann}
	S(\sigma)=0.571241.
\end{equation}
The orthogonalization procedure yields the basis ${\cal B}_{\cal V}=\{|\omega_i\rangle\}$ with
\begin{eqnarray}
	|\omega_1\rangle&=& (0.5, 0.5, 0.5, 0.5)\\
	|\omega_2\rangle&=& (-0.288675, 0.866025, -0.288675, -0.288675) \\
	|\omega_3\rangle&=& (0.408248, 0, 0.408248, -0.816497) \\
	|\omega_4\rangle&=& (0.707107, 0, -0.707107, 0).
\end{eqnarray}
Let the quantum channel be binary, i.e. let $k=2$. The codewords are constructed along $|c(\omega_i)\rangle=| Z_2(i-1)\rangle$, yielding the variable-length states
\begin{eqnarray}
	|c(\omega_1)\rangle&=&|\o\rangle\\
	|c(\omega_2)\rangle&=&|1\rangle\\
	|c(\omega_3)\rangle&=&|10\rangle\\
	|c(\omega_4)\rangle&=&|11\rangle,
\end{eqnarray}
that span the code space ${\cal C}$.
In a neutral-prefix code they are realized by the 2-qbit states
\begin{eqnarray}
	|\tilde c(\omega_1)\rangle&=&|00\rangle\\
	|\tilde c(\omega_2)\rangle&=&|01\rangle\\
	|\tilde c(\omega_3)\rangle&=&|10\rangle\\
	|\tilde c(\omega_4)\rangle&=&|11\rangle\\
\end{eqnarray}	
that span the operational code space $\tilde {\cal C}$,
which is a subspace of the physical space ${\cal R}={\mathbbm C}^2\otimes{\mathbbm C}^2$.
Alice realizes the encoder $C:{\cal V}\rightarrow\tilde{\cal C}$, $C=\sum_i |\tilde c(\omega_i)\rangle\langle\omega_i|$, given by
\begin{equation}
	C=\begin{pmatrix}
	0.5 & 0.5 & 0.5 & 0.5 \\ 
	-0.288675 & 0.866025 & -0.288675 & -0.288675 \\ 
  	0.408248 & 0 & 0.408248 & -0.816497 \\ 
   0.707107 & 0 & -0.707107 & 0 \\ 
	\end{pmatrix}
\end{equation}
and the decoder $D=C^{-1}$, given by
\begin{equation}
	D=\begin{pmatrix}
	0.5 & 0.408248 & -0.288675 & 0.707107 \\ 0.5 & 0 & 0.866025 & 0 \\ 
   0.5 & 0.408248 & -0.288675 & -0.707107 \\ 
   0.5 & -0.816497 & -0.288675 & 0 \\
	\end{pmatrix}
\end{equation}
by gate arrays and gives the decoder to Bob. 
The encoded alphabet is obtained by $|c(x)\rangle=C|x\rangle$.
Alice writes the base lengths of the codewords
\begin{eqnarray}
	\underline L_c(a)=0,\,\underline L_c(b)=\underline L_c(c)
	=\underline L_c(d)=1,\\
	\underline L_c(e)=\ldots=\underline L_c(j)=2
\end{eqnarray}
in a table and calculates the corresponding probabilities
\begin{eqnarray}
	p_0=0.6,\quad p_1=0.3,\quad p_2=0.1
\end{eqnarray}
She constructs Huffman codewords for each length
\begin{equation}
	c_0=1,\quad c_1=01,\quad c_2=00,
\end{equation}
such that the average bit length is
\begin{equation}
	 L'=\sum_{l=0}^2 p_l\cdot l = 1.4,
\end{equation}
which is the optimal value next to the Shannon entropy of the length ensemble
\begin{equation}
	I'=-\sum_{l=0}^2 p_l\,\log_2 p_l=1.29546\ .
\end{equation}
Alice hands the table with the Huffman codewords to Bob and tells him that he must listen to the classical channel, decode the arriving Huffman codewords into numbers, receive packages of qbits, whose size corresponds to the decoded numbers, and add to each package enough leading qbits in the state $|0\rangle$ to end up with 2 qbits. Then he must apply the decoder $D$ to each extended package and he will get Alice's original messages.

Say, Alice wants to send the message $|a\rangle$. She prepares $|a\rangle$ and applies the encoder $C$ to obtain the codeword $|00\rangle$. She looks up the corresponding base length $\underline L_c(a)=0$ and truncates the codeword to $\underline L_c(a)=0$ qbits. In this case there are no qbits left at all, so she sends nothing through the quantum channel and the Huffman codeword for ``length 0'' through the classical channel. Bob receives the classical length information ``0'' and knows that nothing comes through the quantum channel and that in this case he has to prepare 2 qbits in the state $|00\rangle$. He applies the decoder $D$ and obtains Alice's original message $|a\rangle$. In order to send message $|b\rangle$, Alice truncates the codeword to $\underline L_c(b)=1$ qbit and obtains $\frac1{\sqrt2}(|0\rangle+|1\rangle)$. She sends the qbit through the quantum channel together with the classical signal ``length 1''. Bob receives the length message and knows that he has to take the next qbit from the quantum channel and that he has to add 1 leading qbit in the state $|0\rangle$. He applies $D$ and obtains Alice's original message $|b\rangle$.
The whole procedure works {instantaneous} and {without loss of information}. 
We have implemented the above example by a $\text{Mathematica}^{\text{\sffamily TM}}$ program and numerical simulations show that the procedure works fine and the specified compression of quantum data is achieved. (You can find the program and the package at \cite{FelbingerMathematica}). 
\smallskip

Let us look for the compression that has been achieved.

\paragraph{} The quantum code information, i.e. the average number of qbits being sent through the quantum channel,
\begin{equation}
	\underline I_c
	=\sum_{x\in{\cal X}} p(x)\underline L_c(x)=0.5,
\end{equation}
falls below(!) the von-Neumann entropy:
\begin{equation}
	\underline I_c<S=0.571241.
\end{equation}
Such a behaviour has already been suspected in section~\ref{lower}. 

\paragraph{} 
The quantum raw information, i.e. the size of the non-compressed messages, is given by
\begin{equation}
	\underline I_c<I_0=\log_2(\dim{\cal V})=2,
\end{equation}
hence the compression rate on the quantum channel reads
\begin{equation}\label{rc}
	{\rm R}_c=\frac{\underline I_c}{I_0}=0.25.
\end{equation}
In other words, the number of qbits passing through the quantum channel is reduced by 75 \%.
Sending 100 messages without compression requires 200 qbits. Using the compression scheme, Alice typically sends 50 qbits. 

\paragraph{} The sum of both quantum and classical information,
\begin{equation}
	I_{\text{tot}}=\underline I_c+I'=1.79546,
\end{equation}
is smaller than the Shannon entropy~(\ref{shannonsigma}) of the original ensemble $\Sigma$:
\begin{equation}
	I_{\text{tot}}< H=2.02945.
\end{equation}
Thus it is better to use the quantum compression scheme than to simply \emph{tell} Bob on the phone which state he must prepare.
As already suspected, $I_{\text{tot}}$ is still greater than the von-Neumann entropy~(\ref{vonneumann}),
\begin{equation}
	I_{\text{tot}}>S=0.571241.
\end{equation} 
The classical part of the compression depends on the algorithm. Only in the ideal case the information can be compressed down to the Shannon entropy of the length ensemble, given by $I'$. Using the Huffman scheme, the average length $L'=1.4$ represents the information that is effectively sent through the classical channel, such that the total \emph{effective} information is given by
\begin{equation}
	I_{\text{eff}}=\underline I_c+L'=1.9.
\end{equation}

\paragraph{} The the total compression rate of both channels reads 
\begin{equation}
	R_{\text{tot}}=\frac{\underline I_c+I'}{I_0}
	=0.897731<1,
\end{equation}
where it is assumed that the information on the classical channel can be compressed down to its Shannon entropy $I'$. Using the Huffman scheme (as we have done in our example), the information on the classical channel can only be compressed to $L'>I'$, such that the \emph{effective} total compression rate is given by
\begin{equation}
	R_{\text{eff}}=\frac{\underline I_c+L'}{I_0}
	=0.95<1.
\end{equation}
Thus in any case there is an overall compression. For higher dimensional source spaces (hence more letters), the compression is expected to get better (provided the letter distribution is not too uniform). However, the numerical effort for higher dimensional letter spaces increases very fast and we want to keep the example as simple as possible.

\section{Concluding remarks}

We have developped a general framework for variable-length quantum messages and defined an observable measuring the quantum information content of individual states by the number of qbits needed to represent the state by a given code. We derived some basic statements about lossless compression. In particular, we have demonstrated that a quantum message can only be compressed without loss of information if the source message is \emph{a priori} known to the sender. On these grounds, we have worked out a lossless and instantaneous quantum data compression protocol. One can object that there is no use in compressing quantum states that are already \emph{known} to the sender, because then Alice could as well tell Bob classically which of the quantum states she wants to communicate. However, such a pure classical communication would require Bob to have a list of possible messages Alice may send. Moreover, for \emph{arbitrary} quantum messages from the source space, Alice would need infinitely many bits to communicate them through a classical channel to Bob. In contrast to that, in our communication scheme Alice \emph{can} send arbitrary messages from the source message space, but she must know which message she is going to send to get the base length. Bob needs only the decoder and the user instructions for the classical channel, then he can reobtain Alice's original messages with perfect fidelity. The protocol can individually be adapted to a given message ensemble, such that compression is achieved for that ensemble.

\section{Open questions}

It would be satisfying to find an \emph{optimal} compressing lossless quantum code with a tight upper bound related to the von-Neumann entropy. This would represent a quantum analogue to Shannon's relation~(\ref{shannonlossless}).
There might be interesting applications to quantum cryptography. By combining the methods of quantum cryptography with the methods of lossless compression, the efficiency of secure data transfer may possibly be increased.
Furthermore, it would be interesting to see how the framework of variable-length messages applies to quantum computation, since the data stored in the register of a quantum computer could also be regarded as a variable-length quantum message. One could also think about variable-length quantum error-correcting codes.
We hope that the presented work stimulates some more discussion and theoretical research on variable-length quantum coding and its applications.

\section{Acknowledgements}

We like to thank Martin Wilkens, Jens Eisert and Alexander Albus for inspiring discussions and supervisory advice. This work is supported by the EU project EQUIP, the \emph{International Max Planck Research School} IMPRS and the \emph{Deutsche Forschungsgemeinschaft} DFG.


\end{multicols}
\end{document}